# Orientation-dependent chemistry and band-bending of Ti on polar ZnO surfaces

P. Borghetti,[1] Y. Mouchaal,[1,2,*] Z. Dai,[1] G. Cabailh,[1] S. Chenot,[1] R. Lazzari,[1] and J. Jupille[1]

[1] Sorbonne Universités, CNRS-UMR 7588, UPMC Univ Paris 06, Institut des NanoSciences de Paris, F-75005, Paris, France

[2] Laboratoire de Physique des Couches Minces et Matériaux pour l'Electronique (LPCMME), Université d'Oran1 Ahmed Ben bella, BP 1524 EL M'Naouer 31000, Oran, Algeria

* Current address: Université Mustapha Stambouli de Mascara, Faculté des Sciences et de la Technologie, BP 763, Mascara 29000, Algeria

E-mail: borghetti@insp.jussieu.fr

**Abstract**

Orientation-dependent reactivity and band-bending are evidenced upon Ti deposition (1-10 Å) on the polar ZnO(0001)-Zn and ZnO(000$\bar{1}$)-O surfaces. At the onset of the Ti deposition, a downward band-bending was observed on ZnO(000$\bar{1}$)-O while no change occurred on ZnO(0001)-Zn. Combining this with the photoemission analysis of the Ti $2p$ core level and Zn $L_3(L_2)M_{45}M_{45}$ Auger transition, it is established that the Ti/ZnO reaction is of the form Ti + 2 ZnO → $TiO_2$ + 2 Zn on ZnO(0001)-Zn and Ti + $y$ ZnO → $TiZn_xO_y$ + $(y-x)$ Zn on ZnO(000$\bar{1}$)-O. Consistently, upon annealing thicker Ti adlayers, the metallic zinc is removed to leave ZnO(0001)-Zn surfaces covered with $TiO_2$-like phase and ZnO(000$\bar{1}$)-O surfaces covered with a defined (Ti, Zn, O) compound. Finally, a difference in the activation temperature between the O-terminated (500 K) and Zn-terminated (700 K) surfaces is observed, which is tentatively explained by different electric fields in the space charge layer at ZnO surfaces.

# 1. Introduction

Mastering electrical contact of zinc oxide with metals[1,2] is of prime interest in all the actual or potential uses of this semiconductor in optoelectronics and microelectronics[3-5]. Ohmic or Schottky contact are required depending on the type of device that is foreseen. Despite studies based on well-defined single crystals, the control of the barrier height at the metal/ZnO interface is far from being fully rationalized as it depends heavily on (i) surface orientation and polarity, (ii) metal reactivity and deposition method and (iii) annealing conditions. The barrier does not obey the ideal contact which should only be dominated by the affinity of the oxide and the work function of the metal[1,2].

Probably due to the pinning of the Fermi level by interface or defect states, unreactive metal such as noble metals show a rectifying behaviour with a barrier height centred around 0.75 eV independently of the metal work function. However, as a rule of thumb, the contact can be switched from Schottky to ohmic with reactivemetals, the oxides of which present a higher enthalpy[1,2]. Among the potential candidates that were explored, Ti is the one on which attention has been most focused either as deposited or as an alloying element of the contact layer[1,2,6-12]. Ti forms a very low barrier with ZnO both because of its low work function compared to electronic affinity of ZnO and of its high reactivity. It is believed that the interfacial reaction leads to a reduced ZnO subsurface enriched in O-vacancies which act as donors and increase the carrier concentration[1,2,6]. Next to its use to create ohmic contacts[1,2,6-12], titanium is also known to promote adhesion for noble metals within optical coatings deposited on multilayeredglazings[13-14], to enhance the gas sensor properties of ZnO[15] and to set up resistive random access memories[16]. Changes in Ti/ZnO stacking as a function of temperature are crucial regarding the optimization of contact resistivity[1,2,6-12], gas detection limit[17] and mechanical properties of coatings[13]. In all these fields, the interface chemistry is at the heart of the problem. But the detailed mechanism of interface reactivity and the role of the surface orientation in terms of species profile and chemical state are not yet resolved.

Regarding surface science approach of metal/polar-ZnO surfaces, little has been made to address the orientation-dependent chemistry and charge transfer at a reactive metal/ZnO interface. Most studies involving the polar ZnO(0001) and ZnO(000$\bar{1}$) surfaces have focused either on noble metals or on the end of the transition series: Pt[18–22], Pd[23], Cu[24–30], Ag[23,31,32], Au[33]. But dramatic transformations can occur with metals from the beginning and middle of the transition series, as illustrated by the formation of spinels upon annealing Cr/ZnO(000$\bar{1}$)[34,35] and Fe/ZnO(0001)[36] films and ZnO/Ti/ZnO coatings[13] ($ZnCr_2O_4$, $ZnFe_2O_4$ and $Zn_2TiO_4$, respectively). Regarding Ti, the chemical reactions at the interface of sputter-deposited ZnO-Ti multilayers have been explored by hard x-ray photoelectron spectroscopy, revealing the oxidation of Ti and the reduction of ZnO[13,37]. But the study was limited to (0001) oriented sputter-deposited polycrystalline ZnO films with no defined surface polarity[37] which blurs the actual role of the termination. The orientation-dependent chemistry is expected to affect the reaction and inter-diffusion at the metal/ZnO interface. Indeed, the mechanisms of polarity healing are different between the two faces. In ultra-high vacuum conditions, the Zn-terminated ZnO(0001) is stabilized by triangular features[38] providing the required compensation charges through Zn vacancies. On the O-terminated ZnO(000$\bar{1}$), hydroxyl groups play an unmistakable role[39] in polarity healing but the sub-stoichiometric reconstructions have recently been put forward[40,41].

At this stage, three sets of questions arise concerning: (i) the mechanism of the formation of the interface, which requires monitoring the growth from the early stages of the deposition, (ii) the chemical structure of interfaces, and (iii) the way the orientation affects properties. In the present work, the growth of ultrathin Ti films on the basal O-terminated ZnO(000$\overline{1}$) and Zn-terminated ZnO(0001) single crystal surfaces (referred as Zn-ZnO and O-ZnO hereafter) is studied in a ultra-high vacuum chamber by X-ray photoemission spectroscopy (XPS) by scrutinizing the evolution of the Ti $2p$ core level and the Zn $L_3M_{45}M_{45}$ transition. Both the orientationdependent chemistry and band-bending induced by Ti films and their behaviour upon annealing treatment are analysed. Angle dependent emission measurements also allow the determination of the depth profiles of chemical states.

## 2. Experimental Section

Experiments were carried out in a ultra-high vacuum chamber (base pressure $3 \times 10^{-8}$ Pa) equipped with X-ray photoelectron spectrometer (XPS) and low energy electron diffraction device (LEED). Substrates provided by Tokyo Dempa were chemicomechanically polished and post-annealed above 1500 K; they had a resistivity around a few ohms with a low content of Li impurities[42]. O-ZnO and Zn-ZnO surfaces were prepared by cycles of $Ar^+$-ion sputtering (kinetic energy 800 eV) followed by vacuum annealing (T = 1100 K and T = 1300 K, respectively). The absence of surface contaminants was checked at the level of XPS sensitivity (a few *%* of monolayer) while the surface crystallinity was indicated by sharp LEED (1×1) patterns[39]. Ti was deposited by electron beam evaporation from a Ti rod in ultra-high vacuum on ZnO substrates held at room temperature. The absolute deposition rate was calibrated using a quartz microbalance while keeping the evaporation flux constant. Thermal annealing of films was performed by electron-bombarding the back side of the Mo holder on which the ZnO samples were mounted. The annealing temperature was measured on the Mo plate by an optical pyrometer. XPS spectra were obtained using a non-monochromatic Al K₍ source photon energy of h₍ = 1486.7 eV) and an Omicron EA 125 hemispherical analyzer at a pass energy of 20 V.

Electron collection was either normal to the surface or grazing (take off angle 60°) to enhance the surface sensitivity of the measurement. If partial contamination by residual vacuum can never be fully ruled out, ageing of thick Ti films and oxygen uptake in 20h turned out to be moderate. The fitting of Ti $2p$ XPS spectra was done with CasaXPS software using a Shirley background[43], a set of Voigt functions for the $Ti^{4+}$, $Ti^{3+}$, $Ti^{2+}$ oxidation states and a Doniach-Sunjic line profile for the metallic $Ti^0$ component. More details on the spectra fitting are given in the Electronic Supplementary Information (ESI).

## 3. Reduction of ZnO at the interface with a Ti overlayer

The O-ZnO and Zn-ZnO surfaces were exposed to identical stepwise Ti depositions of nominal thicknesses $d_{nom}$ = 1 Å, 2 Å, 4 Å and 10 Å. The Ti deposition had dramatic effects on both surfaces. The total disappearance of the LEED patterns from the initial stages of the film growth ($d_{nom}$ = 1 Å) indicates the loss of the ZnO surface crystallinity and the lack of long-range order for the Ti films. In parallel, the XPS Ti *2p* spectra recorded at normal emission demonstrate the high reactivity of Ti on both polar surfaces (Fig. 1a). According to the binding energy (BE) position of the peaks (vertical dotted lines in Fig. 1a, more details can be seen in ESI), the dominant species at the lowest coverage is $Ti^{4+}$. Upon increasing the Ti thickeness, the lower Ti oxidation states (from $Ti^{3+}$ to $Ti^0$) progressively appear and, starting from $d_{nom}$ = 4 Å, the $Ti^0$ dominates the total composition of the Ti adlayer. The areas of the various Ti components obtained from the spectra fits (see ESI) have been plotted as a function of the total Ti *2p* area in Fig. 1b, which allows the definition of trends specific to each polar surface. For the same $d_{nom}$, both the total Ti *2p* area and the Ti/Zn area ratio (Fig. S1) are larger on O-ZnO than on Zn-ZnO. Such a tendency does not depend on the evaporation sequence, since it is also observed when, instead of stepwise depositions, 2, 4, 10 Å Ti are deposited at once. In passing, this demonstrates the lack of strong effect of residual vacuum at the time scale of the measurements. Assuming the same sticking coefficient on both surfaces, the observation indicates a better Ti wetting on O-ZnO than on Zn-ZnO, a trend which has already been observed for unreactive Ag[31,32] and Cu[26,28]. Additionally, the value of the area of the $Ti^{4+}$ component, which remains constant for all film thicknesses under study, is significantly higher on Zn-ZnO than on O-ZnO (∼25%). By contrast, the area of $Ti^0$ tends to be slightly higher on O-ZnO (Fig. 1b). The interaction of Ti with ZnO is likely complex. As demonstrated afterwards, the presence of all the Ti oxidation states and of metallic Zn on both surfaces as well as the absence of LEED patterns indicate that a partial etching of the ZnO surface occurs on both Zn-ZnO and O-ZnO surfaces. The presence of hydroxyl groups and the specific reconstruction of each polar surface[38–41] are expected to provide various local environments for Ti atoms. Nevertheless, the XPS spectra show clear orientation-dependent trends that are analyzed in what follows.

On both ZnO surfaces, the oxidation of Ti at the interface is accompanied by the reduction of Zn. In the absence of sizeable core-level shift associated to the chemical state of zinc, Zn oxidation is efficiently tracked by means of the Zn $L_3M_{45}M_{45}$ Auger line, the most intense Zn LMM transition [44–46]. In this transition, the core-hole in the Zn $2p_{3/2}$ level ($L_3$) is filled by an electron decaying from the Zn 3d level ($M_{45}$), and the difference in energy between those two levels is used to emit an electron from the Zn 3d level ($M_{45}$). The Zn $L_3M_{45}M_{45}$ profile is dominated by two intense features separated in energy by ∼ 3 eV (peaks A and B in Fig. 2a). The oxidation of the metal results in a shift towards lower kinetic energy (KE) (or higher BE), a broadening of the Zn LMM profile and a change in A/B ratio[44–46]. Peak A of pristine ZnO ($A_{ox}$) and peak B of metallic Zn ($B_{met}$) are separated in energy by ∼ 8 eV, which allows the identification and quantification of the oxidized states of Zn by means of templates recorded on ZnO surfaces and thick metallic Zn films[44–46]. In Fig. 2, Zn $L_3M_{45}M_{45}$ spectra are decomposed into ZnO and Zn contributions by fitting with a linear combination of the above templates. Inspection of the Auger profile reveals (i) that for both ZnO surfaces, the relative intensity of the metallic component $B_{met}$ grows as the Ti coverage increases and (ii) that, in contrast to spectra of the Ti/Zn-ZnO surface that are perfectly decomposed into ZnO and Zn components (Fig. 2b), spectra recorded on Ti/O-ZnO cannot be fitted by such a linear combination (Fig. 2c). Significant extra-contributions centered at BE ∼ 496 eV and BE ∼ 493 eV are present (shaded areas in Fig. 2a and 2c). These are assigned to the formation of an additional Zn species, which is assumed to be related to the formation of a (Ti, Zn, O) compound, herein labelled as TZO, having an intermediate Ti chemical shift. The decrease in intensity of the relative area of this extra-contribution upon increasing Ti thickness indicates that the corresponding Zn species lies at the Ti/ZnO interface.

## 4. Band-bending as a reporter of the Ti/ZnO interaction at the onset of growth

The deposition of Ti on the two polar orientations of ZnO produces quite different changes in band-bending (BB). Fig. 3 shows the energy position of the Zn $2p_{3/2}$ and O $1s$ peaks (Fig. 3a and 3b, respectively) before and after depositing $d_{nom}$ = 1 Å of Ti. The Zn $2p_{3/2}$ and O $1s$ lines recorded on the clean O-ZnO are found at 0.3 eV toward lower BE relative to those observed on Zn-ZnO, which is due to a different space-charge layer formed at the two polar surfaces[47,48]. On O-ZnO, the Ti deposition induces a sizeable BE shift on the Zn $2p_{3/2}$ and O $1s$ core-levels towards higher binding energies (ΔBE= 0.45 ± 0.05 eV, top Fig. 3), while on Zn-ZnO, no shift is observed (bottom Fig. 3). This points to a Ti-induced downward BB on the O-ZnO surface and an absence of BB on the Zn-ZnO surface. The present trend is further confirmed by the decomposition of Zn $L_3M_{45}M_{45}$ spectra of Fig. 2a. In the case of the linear combination used for fitting the Ti/Zn-ZnO spectrum (bottom part of Fig. 2a), the spectrum of bare Zn-ZnO does not display any energy shift upon the Ti deposition of $d_{nom}$ = 1 Å. In contrast, the best fit of the Ti/O-ZnO spectrum (top part of Fig. 2a) results in an energy shift of the bare O-ZnO by +0.15 eV, i.e. in the same direction as the BB for Zn $2p$. By applying a +0.4 eV shift as observed for the core levels (Fig. 3), the curve fitting degrades, which further confirms the assumption of an extra oxidation state other than that of metallic Zn or ZnO.

Comparable BBs, and somehow orientation-dependent chemistry have already been observed upon atomic H adsorption on the basal ZnO surfaces[49–54]. On Zn-ZnO, Becker and coworkers[50,51] highlighted the instability of this surface upon atomic hydrogen exposure. Whereas the substrate keeps a (1×1) LEED pattern, helium atom scattering (HAS)[50,51] and photoemission[49] evidenced the formation of OH groups through the breaking of the Zn-O back bonds. A prolonged exposure destroyed the periodicity as seen by HAS but not by LEED proving that this surface reduction is limited to the top surface layers. Somehow similar findings were made by near field microscopies and spectroscopic ellipsometry[52–54] but at much higher exposure. These measurements revealed that the Zn-ZnO is easily etched by atomic hydrogen giving rise to a surface roughening. The signatures of metallic Zn and OH are also observed by photoemission[49]. At the opposite, the O-ZnO surface passivates upon the formation of an H-covered surface[49]. From ellipsometry and AFM, no changes
are observed on the O-terminated surface[52,53]. The difference of reactivity of the two surfaces stems from the bonding strength between O-H and Zn-H[55]. Besides the loss of atomic structure, similar considerations can be applied to the case of the Ti adsorption on Zn-ZnO, Ti adatoms interact with O atoms of the second layer of Zn-ZnO by "etching" the Zn atoms of the first layer which then form metallic Zn. The substitution of Zn ions by Ti ions has formally no effect on surface dipoles. Consistently, it does not entail any change in BB (Fig. 3, bottom spectra). On the contrary, on the O-ZnO, Ti adatoms bond directly to surface O atoms without altering the O-Zn stacking, which results in a charge transfer from the Ti atoms to the oxide and causes a downward BB (Fig. 3, top spectra). Quite remarkably, a similar picture was proposed for Pt/ZnO interfaces. At the Pt/Zn-ZnO interface, the observation of metallic Zn points to a partial reduction of the oxide. In contrast, no such reduction occurs for Pt/O-ZnO (while Pt-O bonds are evidenced without any disruption of Zn-O bonds)[22]. In present experiments, a partial substitution of hydroxyls groups that are known to exist on O-ZnO as due to residual background pressure cannot be ruled out[39–41] as it was already shown on alumina surfaces[56].

A main consequence of the above description of the Ti/ZnO interfaces is the orientation-dependent chemistry of Ti on the two ZnO polar surfaces. On Zn-ZnO, the etching of the surface Zn atoms by Ti adatoms suggested by the absence of BB leads to a formal substitution of Zn by Ti via the reaction:

$$Ti + 2\ ZnO \rightarrow TiO_2 + 2\ Zn$$

The exclusive existence of only two states of zinc revealed by the Zn LMM transition (Fig. 2) corroborates this scenario, as well as the degree of oxidation of Ti which, at a coverage of 1 Å, is mostly in the $Ti^{4+}$ form (71%) *vs* 19% and 9% for $Ti^{3+}$ and $Ti^{2+}$, respectively (Fig. 1).

On O-ZnO, the sharp variation in BB provides a framework to understand the main aspects of what is going on at the Ti/ZnO interface. The downward BB observed at the onset of the Ti deposition indicates that Ti adatoms sit on surface O atoms to create dipoles that can be formally represented as $Ti^{\delta+}$-$(O-Zn)^{\delta-}$. Consistently, the extra-components seen in the Zn $L_3M_{45}M_{45}$ Auger spectra reveal that Ti reacts with ZnO to produce a TZO interfacial compound (Fig. 2a and 2c). In such a compound, the sharing of O atoms leads to an average formula $TiZn_xO_y$ in which $y$ is likely greater than $x^{57}$. As a consequence, the reduction of ZnO is expected to parallel the formation of the TZO compound, in a reaction of the form:

$$Ti + y\ ZnO \rightarrow TiZn_xO_y + (y-x)\ Zn$$

For a Ti/O-ZnO coverage of 1 Å, the sum of the contributions $Ti^{3+}$ (28%) and $Ti^{2+}$ (23%) of the Ti *2p* spectra is equivalent to the $Ti^{4+}$ component (48%) (Fig. 1), this is indicative of an average degree of oxidation of titanium of 3.2 which is smaller than 4, in particular smaller than in the case of Ti/Zn-ZnO. The formation of the TZO compound explains the relatively moderate oxidation of the Ti adlayer.

## 5. Orientation-dependent depth profile of chemical states

To understand the behavior of the Ti/ZnO films upon increasing Ti thickness, the analysis of the XPS data recorded at $d_{nom}$ = 4 Å is of particular interest. At this coverage, the relative amount of metallic Ti does not yet completely overcome that of the oxidized Ti species (Fig.1b) so that the depth profile can be described by the comparison of normal and grazing (60° off-normal) emission photoemission spectra (NE and GE, respectively). The vertical depth profile of the Ti and Zn species is explored by comparing the relative intensity of Ti *2p* and Zn $L_2M_{45}M_{45}$ spectra (Fig. 4). The Zn $L_2M_{45}M_{45}$ Auger line involves the same shallow levels as the Zn $L_3M_{45}M_{45}$ though the Ti $2p_{1/2}$ core-hole replaces Ti $2p_{3/2}$. Despite its lower intensity, the Zn $L_2M_{45}M_{45}$ has the decisive advantage to make comparisons between NE and GE easier; its proximity to Ti *2p* when using Al $K_\alpha$ x-rays excitation leads to electrons with similar kinetic energies. Therefore, this BE region has been chosen for the systematic study of depth profiles of the 4 Å thick Ti/ZnO films, both after deposition at 300 K (Fig. 4a) and upon annealing (Fig. 4b and 4c). The Ti *2p* spectra are normalized to the same area for the two emission angles, which allows for the straight comparison of the Ti components as well as the Zn $L_2M_{45}M_{45}$/Ti *2p* area ratio.

On the 4 Å thick Ti/Zn-ZnO film after deposition at 300 K (bottom spectra of Fig. 4a), the profile of the Ti *2p* spectrum does not change from NE to GE. The absence of angular-dependence demonstrates that all Ti species, $Ti^{4+}$, $Ti^{3+}$, $Ti^{2+}$ and $Ti^0$, are uniformly distributed across the Ti layer. In parallel, the strong decrease in the Zn $L_2M_{45}M_{45}$/Ti *2p* area ratio at GE relative to NE, which is mainly due to the decrease in intensity of peak $A_{ox}$, shows that the Zn-ZnO substrate is covered by a partially oxidised Ti adlayer. In addition, the slight increase in intensity of $B_{met}$ indicates that Zn tends to diffuse outward. The situation changes dramatically for Ti/O-ZnO (Fig. 4a, top spectra). At GE, the considerable increase of the spectral weight of $Ti^{4+}$ with respect to that of the other species ($Ti^0$, $Ti^{2+}$ and $Ti^{3+}$) proves that $Ti^{4+}$ lies on top of the Ti film. The Zn $L_2M_{45}M_{45}$/Ti *2p* area ratio also contrasts with the Ti / Zn-ZnO case. It increases at GE, which is due to the relative raise in intensity of the peak $A_{ox}$, while the intensity of $B_{met}$ slightly decreases. The Zn lying at the surface of Ti / O-ZnO is therefore oxidized. Although it is not possible to determine whether it belongs to pure ZnO or to a TZO compound involving $Ti^{4+}$, it is demonstrated that, at the extreme surface, *(i)* the degree of oxidation of Ti increases and that *(ii)* the oxidized Ti is accompanied by oxidized Zn. Interestingly, the decomposition of the Zn $L_3M_{45}M_{45}$ spectra indicates that the area of the extra contribution assigned to TZO (shaded areas in Fig. 2a and 2c) decreases from NE to GE, thus suggesting that the oxidized Zn at surface has not the same chemical environment as in the interfacial TZO compound. The present observations are sketched in Fig. 5a, which represents, in a very simplified way, the depth profiles of the Ti-Zn layers on the two ZnO polar orientations.

## 6. Annealing Ti/ZnO interfaces

For both 4 Å thick Ti/Zn-ZnO and Ti/O-ZnO films, XPS Ti *2p* and Zn $L_2M_{45}M_{45}$ spectra recorded after annealing at different temperatures show a progressive oxidation of Ti into $Ti^{4+}$ (Fig. 4b). In parallel, the O *1s* spectra (Fig. 6) reveal an enrichment in oxygen of the top layer. The oxidation process is therefore related to migration of oxygen towards the surface. Upon increasing the annealing temperature, the component $B_{met}$ disappears, while component $A_{ox}$ slightly increases. The likely reason is the progressive desorption of the metallic Zn. For comparison, the metallic Zn/$Al_2O_3$(*0001*) film is readily removed from the alumina surface below 600 K due to the high vapour pressure of metallic zinc[45,46]. Noteworthy, the extra contribution observed in the Zn $L_2M_{45}M_{45}$ and Zn $L_3M_{45}M_{45}$ spectra upon Ti deposition on O-ZnO (Fig. 2a and 2c) disappears after annealing at 1000 K.

The spectra taken at NE after annealing at 1000 K show that, on both surfaces, the overlayer involves Ti and Zn species at their highest degree of oxidation (Fig. 4b). However, the annealed films of the two terminations differ in the vertical distribution of the Zn and Ti species. The angle-dependent photoemission spectra of the samples annealed at 1000 K are shown in Fig. 4c. On Zn-ZnO, the Ti *2p* profile does not change from NE to GE, while the Ti *2p*/Zn $L_2M_{45}M_{45}$ intensity ratio strongly increases. Thus the probed Ti depth is mostly composed by $Ti^{4+}$ ($TiO_2$), while the oxidized Zn remains confined to the substrate below. On OZnO, neither the Ti *2p* profile nor the Ti *2p*/Zn $L_2M_{45}M_{45}$ ratio change from NE to GE which reveals the occurrence of a defined TZO compound; the perfect overlapping NE and GE spectra implies that the sole ZnO substrate is never detected. Thus, Ti has diffused into the ZnO substrate over the depth probed by XPS at NE, which is around 1-2 nm for KE = 1020 eV. In other words, the TZO formation extends well beyond the nominal Ti thickness of 4 Å and implies interdiffusion of Ti, Zn and O. Notably, the trend observed at the onset of the deposition of Ti, i.e. the prevalence of $Ti^{4+}$ on Zn-ZnO and the presence of a TZO compound at O-ZnO, is confirmed on thicker films upon annealing.

At the level of the photoemission, the nature of the formed compound can only be hypothesized; on the basis of the oxidation states of Ti and Zn, Zn ortho-titanate spinel ($Zn_2TiO_4$), Zn polytitanate defective spinel ($Zn_2Ti_3O_8$) and Zn meta-titanate ($ZnTiO_3$, hexagonal ilmenite structure) are potential candidates accordingly to the known compounds and the phase diagram of $TiO_2/ZnO$[6,57,58]. Annealing-induced interdiffusion at the interface of metal oxides is a recurrent mechanism for spinel formation, but the phenomenon is generally studied only at microscopic level[59–61]. The spinel phase has been previously hypothesized on sputter-deposited ZnO/Ti/ZnO coatings annealed at 900 K[13,37] as well as at the Cr/O-ZnO interface and at the Fe/Zn-ZnO interface upon annealing at 830 K[34,35] and 820-870 K[36],respectively. Interestingly, a thin zinc oxide layer above metallic chromium is observed at room temperature for Cr/O-ZnO, in analogy to our results[34,35]. Finally, the activation temperature for the Ti oxidation dramatically differs for the two polar surfaces. On O-ZnO, the reaction onsets below 500 K and is completed at 700 K (Fig. 4b). Conversely, on Zn-ZnO, the oxidation process is activated above 700 K and is almost completed at 1000 K (Fig. 4b). In parallel, on both surfaces, the oxygen enrichment of the film (Fig. 6) is activated for similar temperature ranges as for the Ti oxidation. The formation of the spinel observed only on O-ZnO and not on Zn-ZnO as well as the lower activation temperature may be explained by different space charge layers[47,48] at the two surfaces of ZnO that much favours out-diffusion of oxygen and inwards diffusion of Ti on O-ZnO. Similar arguments have been put forward to explain strong metal support interaction phenomena[62] as well as change in wetting of Cu on ZnO[40].

## 7. Conclusion

Different reactive behaviours of titanium with the two polar faces of ZnO have been observed. At submonolayer Ti coverage, changes in band-bending that compare to those produced by hydrogen adsorption are observed. They stem from the orientation-dependent Ti/ZnO surface chemistry. On Zn-ZnO, Ti reduces ZnO to form a Ti oxide, while on O-ZnO, the deposition of Ti gives rise to the formation of a (Ti, Zn, O) compound. A similar chemistry is observed upon annealing the Ti adlayers, although with very different activation temperatures, 500 K on O-ZnO and 700 K on Zn-ZnO. Those orientation-dependent behaviours are expected to strongly affect applications relying either on rather thin Ti/ZnO films or on surface properties that depend on the charge compensation mechanism at the interface of the adlayer with polar ZnO substrates[63]. The present results also partly explain why Ti/ZnO electrical contact properties are quite scattered and depend on annealing treatments and crystal orientation[1].

## Acknowledgments
This work was supported by grants from the ANR, project COCOTRANS: ANR-2011-RMNP-010-02, from Erasmus Mundus Action 2, Erasmus Mundus Maghreb, No.2012-2623 and from the Chinese Scholarship Council: No.201406150013. P. B. acknowledges financial support by the European Union's Horizon 2020 research and innovation programme under the Marie Sklodowska-Curie grant agreement No.658056.

**Figures**

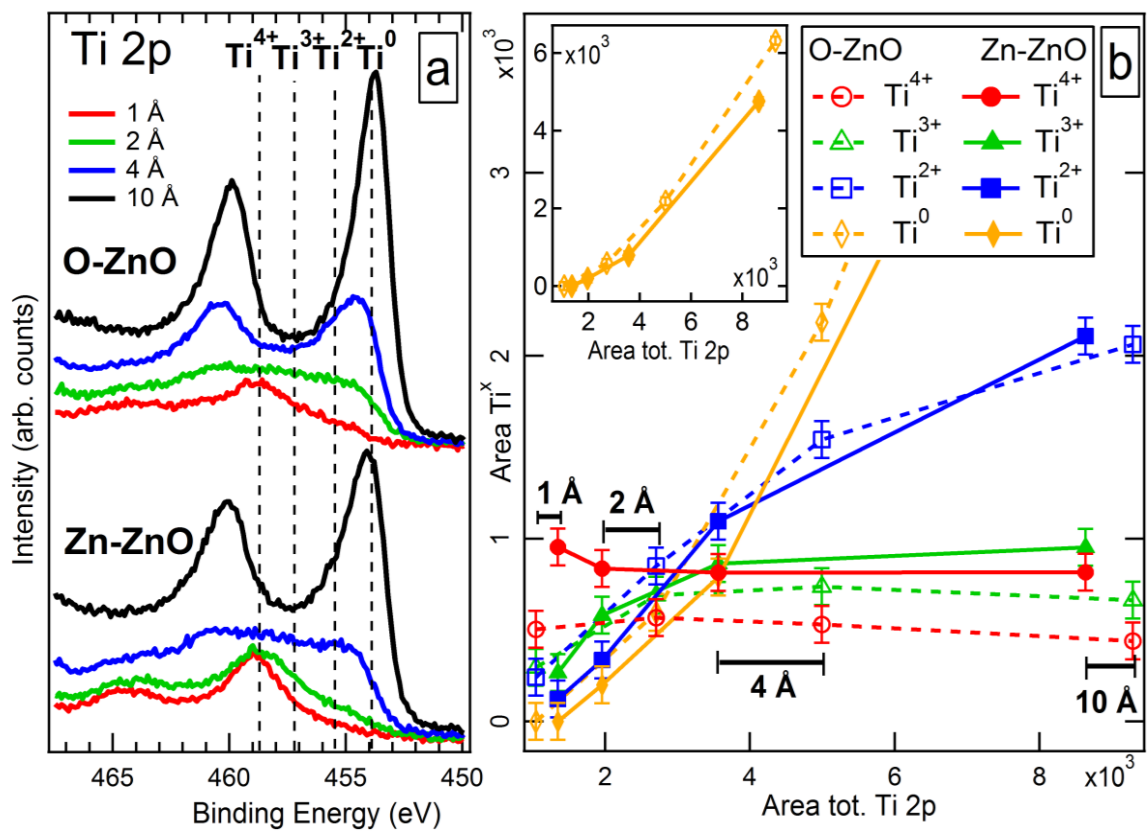

**Fig. 1** XPS analysis of the Ti deposition on ZnO surfaces: a) Ti *2p* spectra on O-ZnO (top) and Zn-ZnO (bottom). Spectra are recorded at a nominal Ti thickness of $d_{nom}$ = 1 Å (red), 2 Å (green), 4 Å (blue) and 10 Å (black) and normalized to the same background on the low BE side (~445 eV). The dotted vertical bars mark the BE of $Ti^{4+}$, $Ti^{3+}$, $Ti^{2+}$ and $Ti^{0}$ as deduced from the fit of the Ti *2p* spectrum of $d_{nom}$ = 1 Å on the Zn-ZnO (see ESI); b) Areas of $Ti^{4+}$ (red), $Ti^{3+}$ (green), $Ti^{2+}$ (blue) and $Ti^{0}$ (orange) components as a function of the Ti *2p* total area recorded on O-ZnO (empty markers, dotted line) and Zn-ZnO (full markers, solid line) films. The corresponding nominal Ti thickness is indicated by the black horizontal bars. The full range of $Ti^{0}$ values is reproduced in the inset. The O-ZnO and Zn-ZnO surfaces were exposed to identical stepwise Ti depositions of nominal thicknesses $d_{nom}$ = 1 Å, 2 Å, 4 Å and 10 Å. The Ti deposition had dramatic effects on both surfaces. The total disappearance of the LEED patterns from the initial stages of the film growth ($d_{nom}$ = 1 Å) indicates the loss of the ZnO surface crystallinity and the lack of long-range order for the Ti films. In parallel, the XPS Ti *2p* spectra recorded at normal emission demonstrate the high reactivity of Ti on both polar surfaces (Fig. 1a). According to the binding energy (BE) position of the peaks (vertical dotted lines in Fig. 1a, more details can be seen in ESI), the dominant species at the lowest coverage is $Ti^{4+}$. Upon increasing the Ti thickness, the lower Ti oxidation states (from $Ti^{3+}$ to $Ti^{0}$) progressively appear and, starting from $d_{nom}$ = 4 Å, the $Ti^{0}$ dominates the total composition of the Ti.

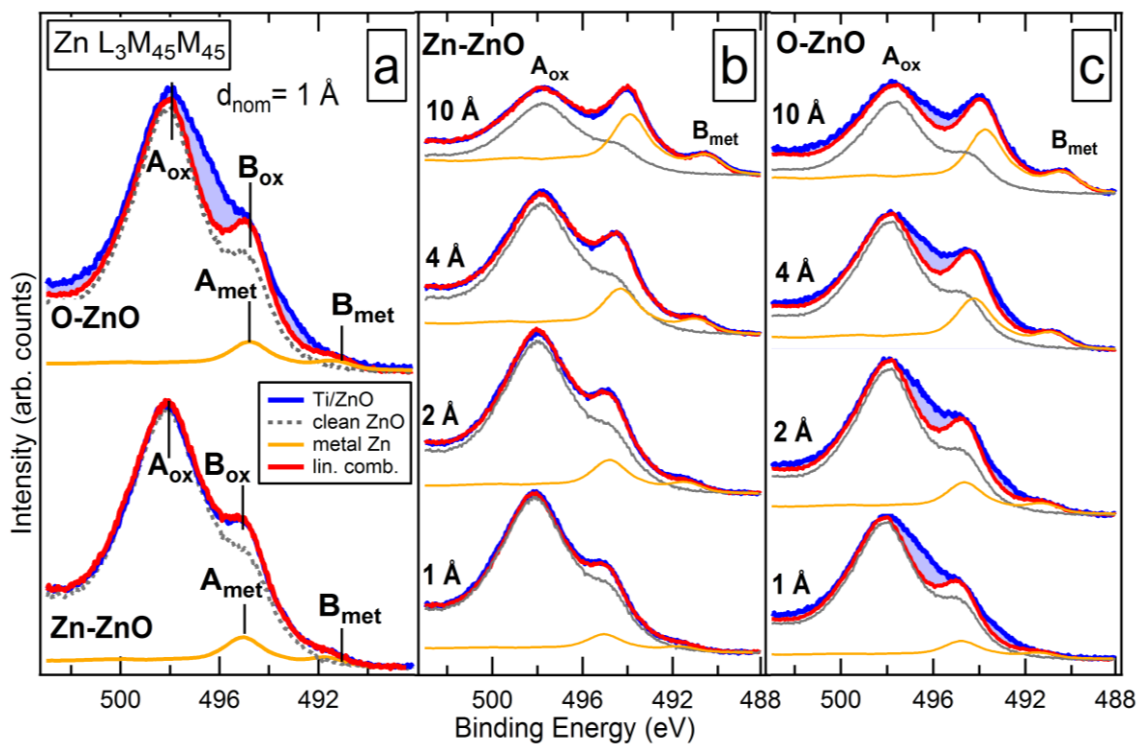

**Fig. 2** Fit of Zn $L_3M_{45}M_{45}$ spectra (blue line) with a linear combination (red line) of the spectra of clean ZnO (dotted gray line) and metallic Zn (orange line): a) spectra obtained by depositing $d_{nom}$ = 1 Å of Ti on O-ZnO (top) and on Zn-ZnO (bottom); b) set of spectra recorded for different Ti coverages (given in figure) on Zn-ZnO and c) O-ZnO. In O-ZnO spectra, extra contributions that are not accounted for by the linear combination are marked by shaded areas.

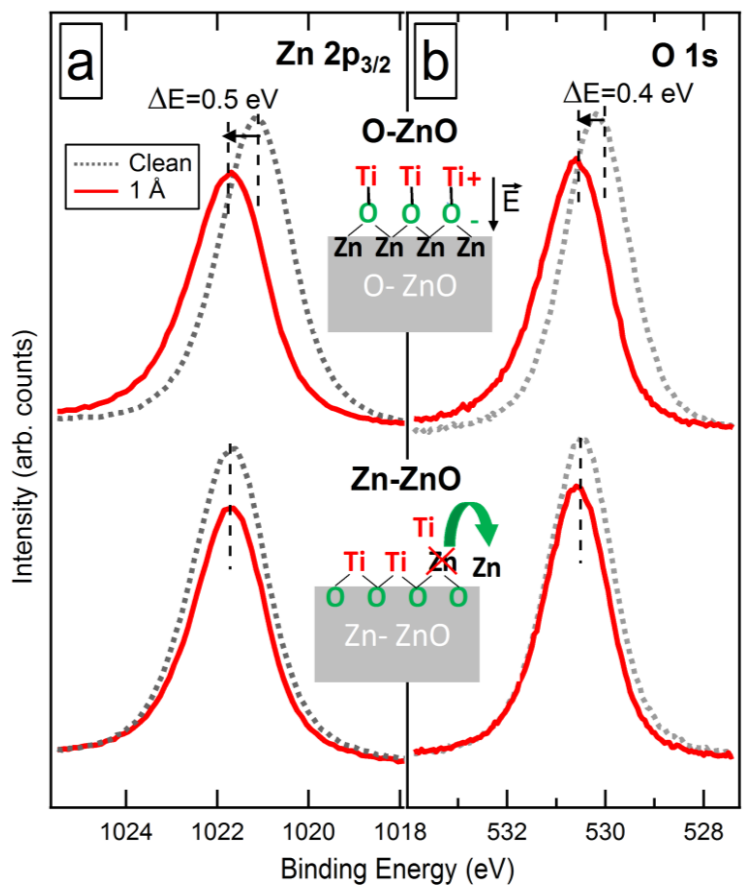

**Fig. 3** XPS spectra of a) Zn *2p*$_{3/2}$ and b) O *1s* peaks before (grey lines) and after (red lines) depositing $d_{nom}$ = 1 Å of Ti. Spectra on O-ZnO and on Zn-ZnO are shown in the top and bottom part of the figure, respectively. Ti/ZnO reactions occurring on each termination are schematized in the figure. The peak shift on O-ZnO demonstrates a downward band-bending.

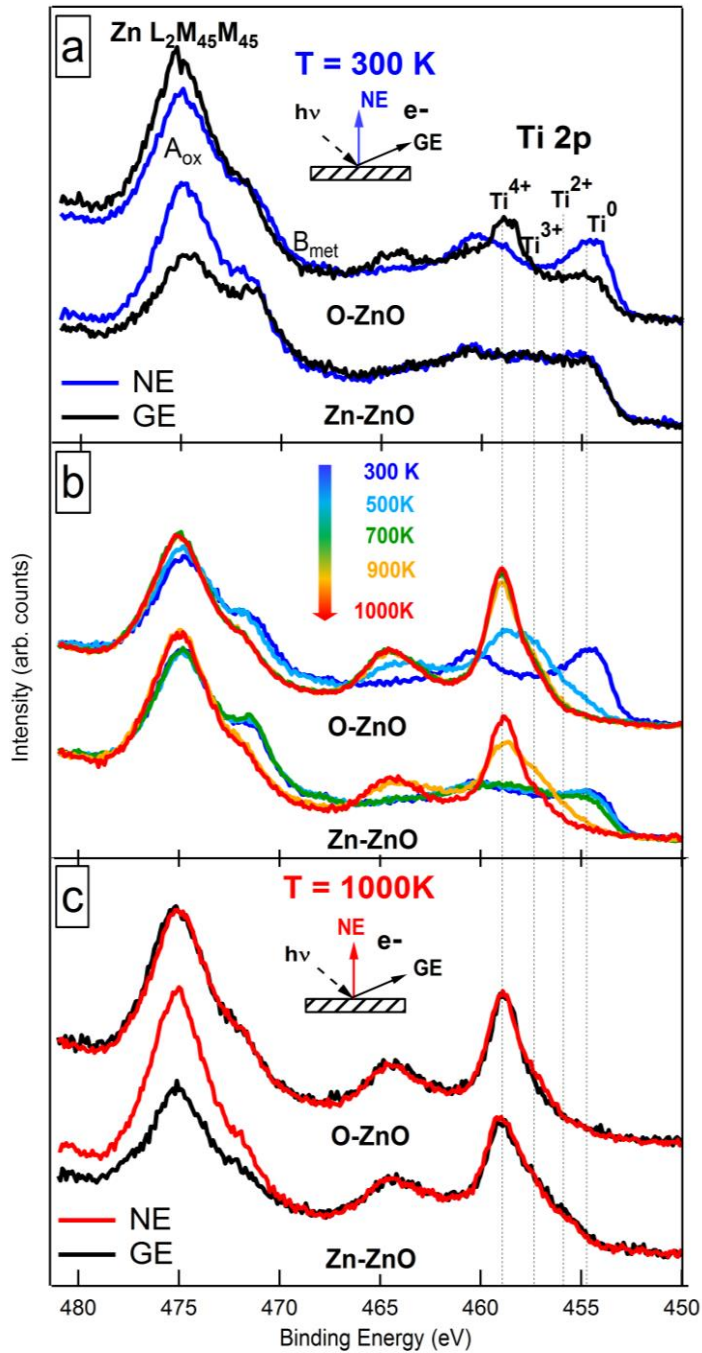

**Fig. 4** Region of Ti *2p* and Zn L$_2$M$_{45}$M$_{45}$ for the thickness d$_{nom}$ = 4 Å of Ti deposited on O-ZnO (top) and on Zn-ZnO (bottom). The spectra are taken at (a) normal emission (NE, blue) and grazing emission (GE, black) after deposition at room temperature, (b) upon annealing (NE) (c) at normal emission (NE, red) and grazing emission (GE, black) after annealing at 1000 K.

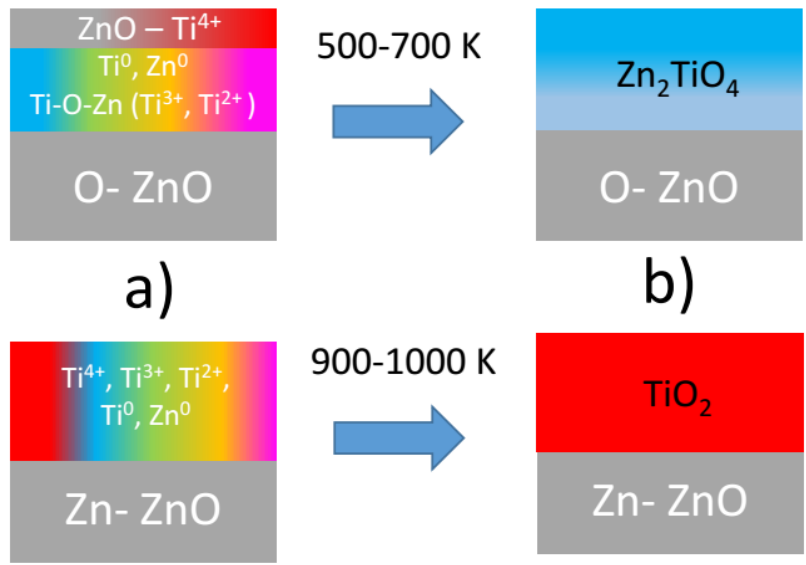

**Fig. 5** Schematic depth profiles of Ti-Zn layer for the thickness $d_{nom}$ = 4 Å of Ti deposited on Zn-ZnO (bottom) and O-ZnO (top) a) before and b) after annealing (see text for explanations).

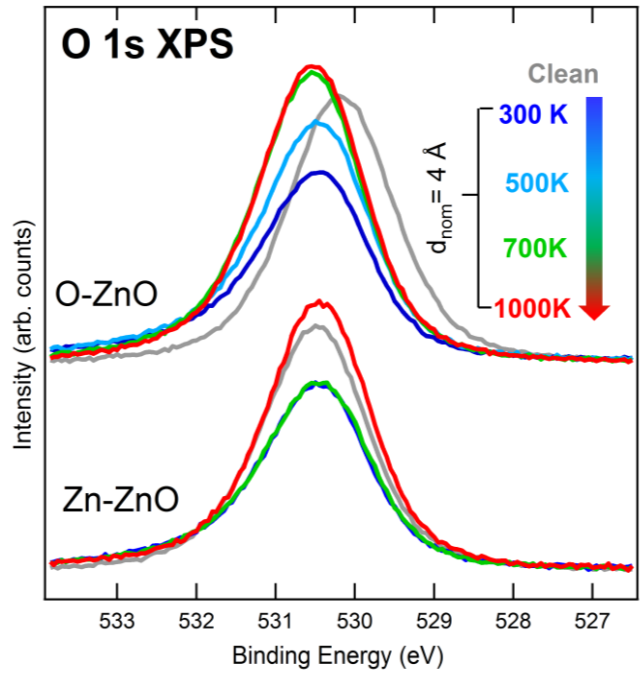

**Fig. 6** O *1s* spectra for the thickness d$_{nom}$ = 4 Å of Ti deposited on O-ZnO (top) and on Zn-ZnO (bottom) upon annealing.